# SOLID STATE QUANTUM COMPUTING USING SPECTRAL HOLES

M.S. Shahriar[1], P.R. Hemmer[2], S. Lloyd[3], J.A. Bowers[1], A.E. Craig[4]

[1] *Research Laboratory of Electronics, Massachusetts Institute of Technology, 77 Massachusetts Avenue, Cambridge, MA 02139*
[2] *Air Force Research Laboratory, Sensors Directorate, Hanscom AFB, MA 01731*
[3] *Department of Mechanical Engineering, Massachusetts Institute of Technology, 77 Massachusetts Avenue, Cambridge, MA 02139*
[4] *The Spectrum Lab, Montana State University – Bozeman, Bozeman, MT 59717*

**A quantum computer that stores information on two-state systems called quantum bits or qubits must be able to address and manipulate individual qubits, to effect coherent interactions between pairs of qubits, and to read out the value of qubits.[1,2] Current methods for addressing qubits are divided up into spatial methods, as when a laser beam is focused on an individual qubit[3,4,5] or spectral methods, as when a nuclear spin in a molecule is addressed using NMR.[6,7] The density of qubits addressable spatially is limited by the wavelength of light, and the number of qubits addressable spectrally is limited by spin linewidths. Here, we propose a method for addressing qubits using a method that combines spatial and spectral selectivity. The result is a design for quantum computation that provides the potential for a density of quantum information storage and processing many orders of magnitude greater than that afforded by ion traps or NMR. Specifically, this method uses an ensemble of spectrally resolved atoms in a spectral holeburning solid. The quantum coupling is provided by strong atom-cavity interaction. Using a thin disc of diamond containing nitrogen-vacancy color centers as an example, we present an explicit model for realizing up to 300 coupled qubits in a single spot. We show how about 100 operations can take place in parallel, yielding close to $4 \times 10^4$ operations before decoherence.**

The basic concept is illustrated in figure 1. Consider a small volume element of a crystal containing a set of impurity atoms. Each atom sees a unique surrounding, so that the resonance frequency for a given transition is different for different atoms. The number of spectrally resolvable bands, $N_R$, is determined by the ratio of the spectral spread to the width of the individual resonance. We consider a situation where the number of atoms in the selected volume is less than $N_R$, so that each atom can be addressed individually.

We choose an effective density low enough to ignore the atom-atom direct coupling. Instead, we couple the atoms in a controlled fashion by placing them in an optical cavity with a strong vacuum Rabi frequency. Once two atoms are coupled by the cavity field, a variety of methods are potentially available for effecting quantum logic between them: essentially any form of coupling between two spectral holes, combined with the ability to perform single-hole quantum operations, allows the implementation of general purpose quantum computations.[8,9] The method we choose is determined by the desire to perform two qubit operations accurately and with a minimum of decoherence. This method is analogous to a scheme proposed by Pellizzari, et al.[4] that uses adiabatic transfer to move quantum information coherently from one particle to another, then performs quantum logic by inducing single-particle Raman transitions.

Consider a situation where each atom has a Λ-type transition, with two non-degenerate spin states coupled to a single optically-excited state, as shown in figure 1. For two atoms separated by a frequency matching the energy difference between the low lying states, choose a cavity frequency that excites a resonance in each atom. Via this common excitation, a cavity photon can act as a 'quantum wire' over which the atoms can exchange optical coherence. Our qubits are stored on spins, however, and so we must use optical coherence to transfer spin coherence. This is accomplished by applying, for each of the two atoms, a laser beam coupling the remaining leg of the Λ transition. The resulting two-photon excitation acts effectively as a cavity mode exciting the spin transition, with the advantage that the excitation can be turned on or off at will by controlling the laser beams. The atoms can use the two-photon mediated quantum wire to exchange spin coherence with each other. If we tune the frequencies of the cavity as well as the laser beams by the amount corresponding to the spin transition, we can couple one of these atoms to a third one. In general, this scheme allows us to produce nearest-neighbor information exchange among a discrete set of N atoms, where N is given by the ratio of the inhomogeneous broadening to the spin transition frequency. Finally, M different spots, spread in two dimensions, can be coupled by using the spatially selective version[4] of this technique, so that in principle up to MN qubits can be all coupled to one another.

In this scheme, each atom has a pair of Λ transitions and six low-lying spin states (figure 2); for illustrative purposes it is convenient to think of these six states as resulting from multiplexing of the spin states of two constituent pseudo-particles, a spin 1 particle (A :purple) and a spin ½ particle (B: turquoise) in each atom. In-between gate operations, the logic states |0> and |1> of the qubit corresponding to every atom are stored in the spin up and down states, respectively, of B, with A in the spin horizontal state, carrying no information. Whenever it is necessary to perform a gate operation between two neighboring gates, the information is first extracted from these storage levels. After this restoration, the logic states |0> and |1> of the qubit corresponding to atom 1 are represented by the spin up and down states, respectively, of A, with B in the spin up state, carrying no information. On the other hand, the logic states |0> and |1> of the qubit corresponding to atom 2 are represented by the spin up and down states, respectively, of B, with A in the spin up state, carrying no information.

To entangle these two atoms, the quantum wire is used first to exchange, for example, the quantum states of the A particles. This results in the atom 2 containing *both bits* of information: 1 in B and 1 in A, unentangled. Quantum logic operations on the two qubits now correspond to simple transitions between the spin sublevels inside atom 2. Such transitions can be used to perform controlled-NOT gates and to entangle B and A. The quantum wire is now used again to exchange the states of A, resulting in the corresponding entanglement of atom 1 and atom 2. By combining inter-atom quantum wires and intra-atom quantum logic gates, it is possible to build up arbitrary quantum logic circuits.

Under circumstances where the ground state splittings are very large compared to the natural linewidth, this method works even when the two Λ-transitions (*a-g-c* and *b-h-d*) are non-degenerate: $\varepsilon_{ac} \neq \varepsilon_{bd}$, but $|\varepsilon_{ac} - \varepsilon_{bd}| \ll \varepsilon_{ac}$. In this case, the cavity is detuned away from resonance; the detuning must be large enough to ignore direct optical excitation, but small compared to $\varepsilon_{ac}$. The laser beam for each atom must have two different frequencies, also detuned so that one of them is two-photon resonant with the *a-g-c* transition, while the other is two-photon resonant with the *b-h-d* transition.

Figure 3 illustrates the steps used in producing an entangled state of the form $\alpha|00\rangle + \beta|11\rangle$, starting from the state $\alpha|00\rangle + \beta|10\rangle$. More general entangled states can be produced using these same steps. Figure 3a shows the process of retrieving the quantum information from the storage levels. The curved arrows represent intra-atomic, Raman π pulses, using laser beams only. Figure 3b shows the steps for producing intra-atomic entanglement. First, a laser-cavity two-photon π pulse for atom 1 is used to transfer the spin coherence from $A_1$ (particle A in atom 1) to the quantum states of the cavity, as a superposition of 0 and 1 photons. A second laser-cavity two-photon π pulse, now for atom 2, transfers this information to $A_2$. In practice, a counterintuitive pulse sequence would be used to effect the same transfer adiabatically, which has the advantage of not suffering from spontaneous emission during the transfer.[4] An intra-atomic Raman π pulse is now used to entangle $A_2$ and $B_2$. Finally, a reverse sequence is used to exchange $A_1$ and $A_2$, producing inter-atomic entanglement. Figure 3c shows the final step of transferring each qubit to its storage level, producing the desired state, which corresponds to a controlled-NOT operation between the two qubits.

The technology that can be used to implement the proposed method is spectral hole burning (SHB).[10,11] In an SHB medium, the number of resolvable lines, $N_R$, can be as high as $10^7$. Here, we present a specific SHB material for implementing this technique: nitrogen-vacancy color center in diamond (NV-diamond).[12] The relevant energy levels and their correspondence to the model are illustrated in figure 4. As can be seen, both degenerate (4a) and non-degenerate (4b) Λ transitions can be realized in this material. While the degenerate case is conceptually simpler, in the case of diamond it has the disadvantage that the hyperfine splittings are small, and comparable to the natural linewidth, which in turns limits the maximum number of gate operations. As such, in what follows we will concentrate on the non-degenerate case. While this system is not perfect, it does meet the basic requirements, and allows us to discuss a concrete model of quantum computing via spectral hole burning. A wide variety of SHB materials exist, including quantum dots, and it may be possible to design media that are optimized for quantum computing.

During the adiabatic transfer step, there are two primary sources of decoherence: dephasing of the spin coherence and cavity losses. For example, the spin decoherence time $T_2$ in diamond is about 0.1 msec. Several NMR spectroscopic techniques exist for increasing this to the $T_1$ limit, and have recently been shown to be applicable to quantum computing.[13] While $T_1$ has yet to be measured for NV-diamond, comparison with other solids (e.g., Pr:YSO, where $T_2$ is about 0.5 msec, while $T_1$ is on the order of 100 sec)[14] suggests that $T_1$ is expected to be of the order of a few seconds. In addition to the NMR type approach, one can use a diamond host free of the $^{13}C$ isotope, which is known to be the limiting source of spin dephasing.

Decoherence due to cavity losses can be minimized by using cavities with long photon lifetimes. Specifically, we have estimated that a concentric, hour-glass cavity, with a mirror separation of L cm, a waist size of D μm, and a Q of $2X10^5$ can be used to achieve, for diamond, a vacuum Rabi frequency of $(170/DL^{1/2})$ MHz, and a cavity width of $(42/L)$ kHZ.[15,16] Choosing D=50 and L=30, we get a Rabi frequency of 0.62 MHz, and a cavity width of 1.4 kHz. The number of operations that can be performed before cavity decay is more than 400. The cavity lifetime ($1/2\pi*1.4$ msec) is close to the $T_2$, so that the cavity still limits the maximum number of operations. The number of operations that can be performed reliably could possibly be extended considerably beyond these numbers by using error-correcting codes developed to circumvent cavity decay under similar circumstances[17,18].

The cavity design mentioned above may enable parallel operation, coupling many different pairs simultaneously. The free spectral range (FSR) of the cavity would be ~250 MHz. By adjusting the value of L slightly, the FSR can be made to be a sub-multiple of the channel spacing, which is about 2.8 GHz. Thus, in principle, all 300 channels could operate simultaneously. However, in order to avoid undesired excitations, at least one channel must be in the storage levels in-between two active pairs. This limits the maximum number of parallel operations $N_p$ to about 100, so that the total number of operations before decoherence can approach $4X10^4$ even without error correction. Of course, one must provide $N_p$ different control beams as well. In principle, this can be achieved as follows: A series of acousto-optic modulators, each operating at 2.8 GHz, will be used to generate $N_p$ sets of control beams from a reference set. The intensities of each set can be controlled independently, and the beams can be combined using a holographic multiplexer, for example[19].

The controlled-not operation described above assumes exactly one atom per spectral channel. To achieve this condition, one can start with a high dopant concentration, in order to ensure that all channels have at least one atom. Consider first the case of NV-diamond. For a 10 μm thick sample, the volume at the waist of the cavity is about V=25000 μm$^3$. For a dopant concentration of $10^{17}$ cm$^{-3}$, the number of atoms is four orders of magnitude bigger than the number of discrete channels. The probability for having at least one atom per channel at an acceptable frequency would therefore be high. The excess atoms in each channel can be removed by a gating process whereby a center can be deformed via excitation at a frequency different from the transitions to be used, and no longer responds to the optical excitation of interest[20]. Irradiation by a much shorter wavelength (e.g., 488 nm) can restore the center to its desired form. Explicitly, the atoms will first be pumped optically into one of the ground-state hyperfine levels. The cavity will then be tuned to the corresponding leg of the Raman transition at a desired frequency. A probe laser pulse tuned to the same transition will experience a shift in the cavity transmission frequency[21] proportional to the number of atoms in this channel. The deforming laser, tuned to this channel, will be pulsed on, while monitoring the cavity frequency-pull. The discrete step-size in the reduction of this frequency-pull will be used to reduce the number of centers to unity. Since this deformation lasts for hours as long as the temperature remains below 4K, there will be virtually no time constraints in preparing all the channels in this fashion. Finally, when considering atom-atom coupling, such as dipole-dipole interaction, the parameter of interest is the mean distance, $V^{1/3}$, between the two atoms in adjacent channels. This distance is about 30 µm, much bigger than the laser wavelength. As such, the atom-atom coupling can be neglected.

To extract information from the qubits, several techniques could be used. For example, a high Q cavity could be used to detect whether an atom is in a particular ground state sublevel by applying an optical π-pulse to an appropriate transition to drive it into an excited state. Once excited, the atom can be probed by a variety of techniques such as frequency pulling of the high Q cavity.[21]

In summary, we have proposed the use of spectral holeburning materials for constructing quantum computers that have the potential to scale up to a large number of qubits. For purposes of illustration, we have explicitly outlined the steps needed to perform a quantum controlled-not using NV-diamond. For this system, upto 300 qubits can be coupled, each to its nearest spectral neighbors, within a single spot. We also point out how about 100 operations can take place simultaneously, yielding close to $4 \times 10^4$ operations before decoherence, even without error correction. This work was supported by AFOSR and ARO.

**Figure Captions:**

**Figure 1:** Schematic Illustration of coupling inhomogeneously broadened atoms using spectral selectivity. The top figure shows a small volume of a crystal, selected by the intersection of the cavity mode and the control laser beams. The bottom figure shows how the atoms can be indexed in terms of their frequency response. Spectrally adjacent atoms, with a frequency difference matching the ground state splitting, can be coupled selectively by tuning the cavity and the coupling lasers. Atom M can be addressed spectrally via the red transition, atom M+1 can be addressed via the green transition, and the two are coupled to the cavity via the blue transition. The key constraint on the matching is that the $\Lambda$ transition in each atom must be two-photon resonant. This can be realized by choosing the laser frequencies appropriately. It is also necessary to make sure that there is only one atom per spectral channel.

**Figure 2:** Relevant energy levels and transitions required of two spectrally adjacent atoms in this scheme. In each atom, the six low-lying levels can be thought of as corresponding to the spin states of two pseudo-particles: a spin 1 particle (A: purple) and a spin ½ particle (B: turquoise ). In the quiescent state, the qubit in each atoms is represented by the spin-up (0=e) and spin-down (1=f) states of B, with A in the spin-horizontal state, containing no information. Whenever it is necessary to perform a gate operation between two neighboring qubits, the qubit in atom 1 ($\alpha_1|e>+\beta_1|f>$) is transferred to the spin-up and -down states of $A_1$, with $B_1$ in the spin-up state ($\alpha_1|a>+\beta_1|c>$). This pattern is alternated in the subsequent atoms in the chain. The qubit in atom 2 ($\alpha_2|e>+\beta_2|f>$) is transferred to the spin-up and -down states of $B_2$, with $A_2$ in the spin-up state ($\alpha_2|a>+\beta_2|b>$). Using a sequence of pulses from the green and red lasers, the quantum states are exchanged between $A_1$ and $A_2$, via the "quantum wire" provided by the blue cavity photon[4]. Conceptually, this can be thought of as a two-step process. First, the red laser transfers the state of $A_1$ to the cavity, producing a superposition of 0 and 1 photons ($\alpha_1|1>+\beta_1|0>$). The green laser then transfers this state to $A_2$ ($\alpha_1|\downarrow>+\beta_1|\uparrow>$). All four bits of information are now in atom 2; as such, any desired gate operation (see fig. 4) can be achieved by a pulse coupling any two of the states (a,b,c,d), using a two-photon transition. $A_2$ is now exchanged with $A_1$ by using a reverse sequence of the red and green lasers. Finally, the state of each atom is transferred to the levels e and f. These storage levels are needed to ensure that the neighboring qubits remain unaffected by these gate operations.

**Figure 3:** Illustration of the steps necessary to produce entanglement, starting from a joint state $(\alpha|0>+\beta|1>)\otimes|0>$. (a) The state of each qubit is retrieved from the storage levels, using off-resonance Raman $\pi$ pulses, producing $(\alpha|a>+\beta|c>)\otimes|a>$. Either polarization selection rules or an external magnetic field can be used to provide the selectivity of the desired transition. (b) A pulse sequence of the red and green lasers exchanges, via the common cavity photon, the states of $A_1$ and $A_2$ (see fig. 2), producing $|c>\otimes(\alpha|c>+\beta|a>)$. Another off-resonance Raman $\pi$ pulse is used to transfer a to b, producing $|c>\otimes(\alpha|c>+\beta|b>)$; this is a controlled-NOT operation that entangles $A_2$ and $B_2$. A reverse pulse sequence of red and green lasers exchanges back the states of $A_1$ and $A_2$, producing $(\alpha(|aa>+\beta|cb>)$, which represents an entangled state of the two atoms. (c) The state of

each qubit is now returned to the storage levels, producing the final state of (α(|00>+β|11>), corresponding to a controlled-NOT operation between the two qubits.

**Figure 4:** Relevant subset of energy levels of the candidate material: N-V color centers in diamond. The $^3A_1$ to $^3E$ transition is excited at 637.8 nm, with a homogeneous linewidth of 5 MHz and an inhomogeneous linewidth of 1 THz at liquid helium temperature. The energy sublevels correspond to the spin orientations of the two uncoupled electrons (S) and the nucleus (I) of the substitutional nitrogen atom. [A] The levels at zero magnetic field. The inter-qubit frequency spacing in this case is 4.6 MHz, corresponding to more than $10^5$ qubits per spot. [B] The levels at a magnetic field of 500 Gauss, including nuclear Zeeman splitting of 300 Hz/Gauss. The inter-qubit spacing in this case is 2.8 GHz, corresponding to about 300 qubits per spot.

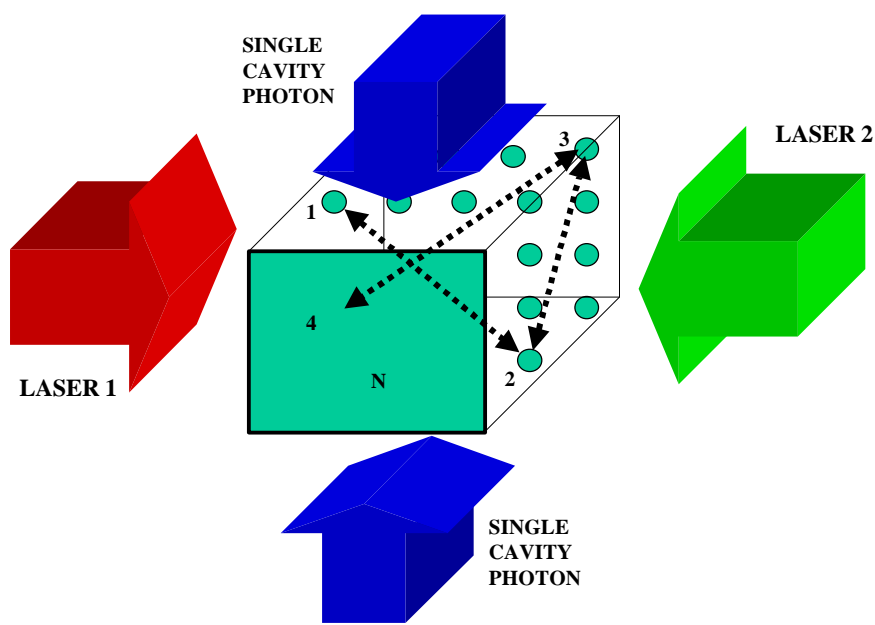

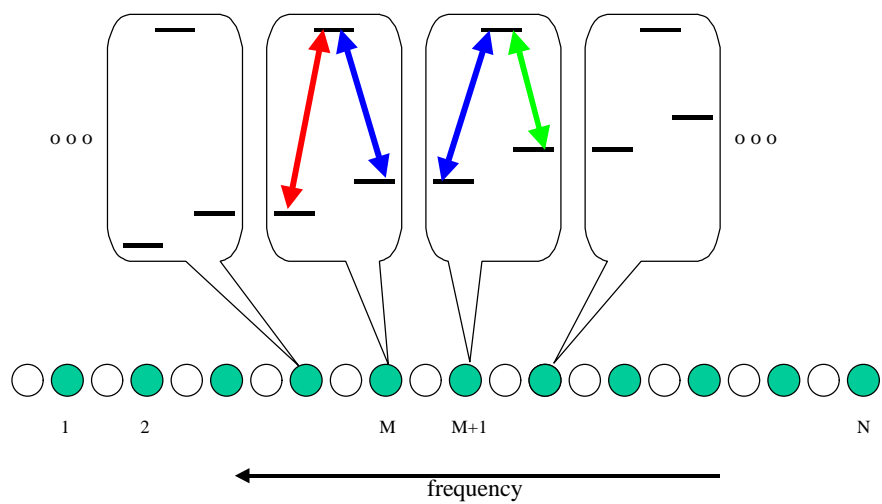

**FIGURE 1**

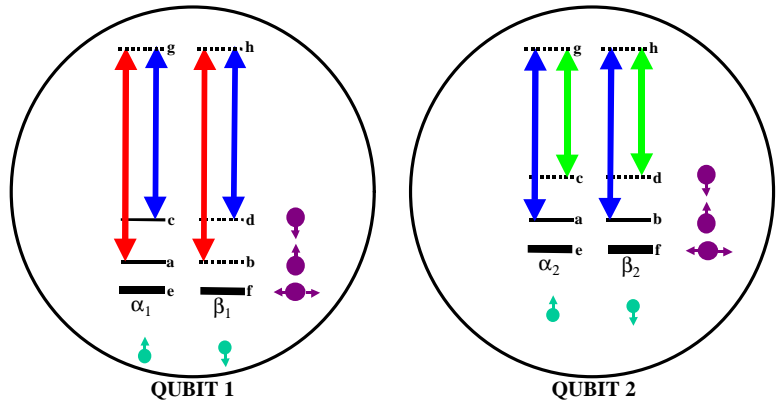

**FIGURE 2**

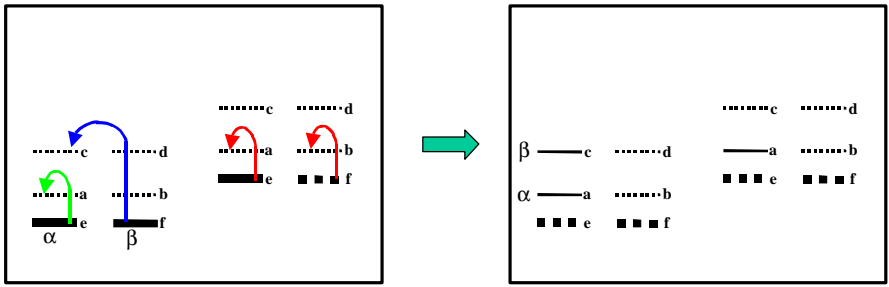

[A]

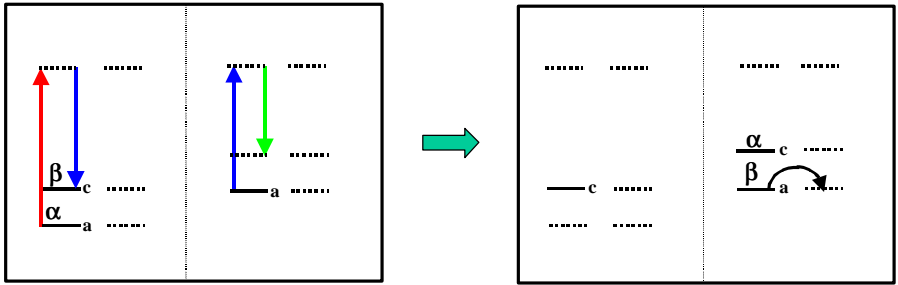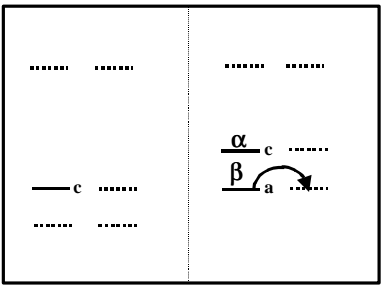

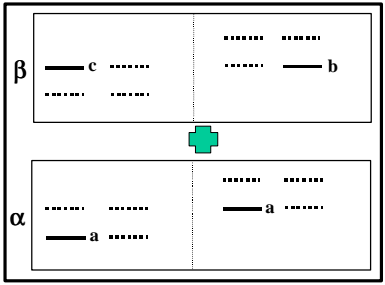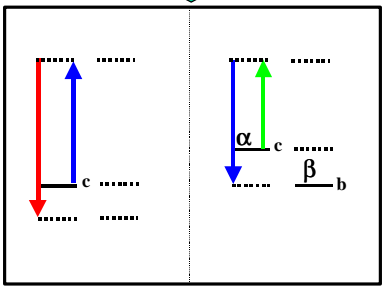

[B]

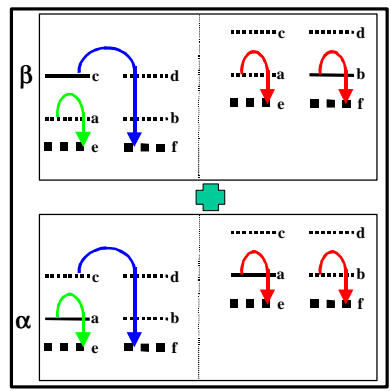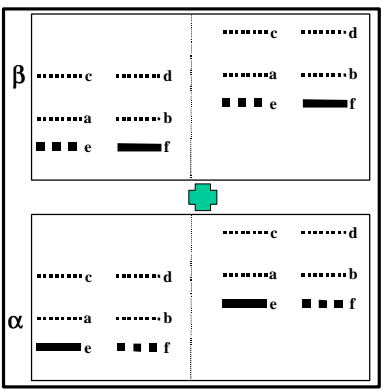

[C]

**FIGURE 3**

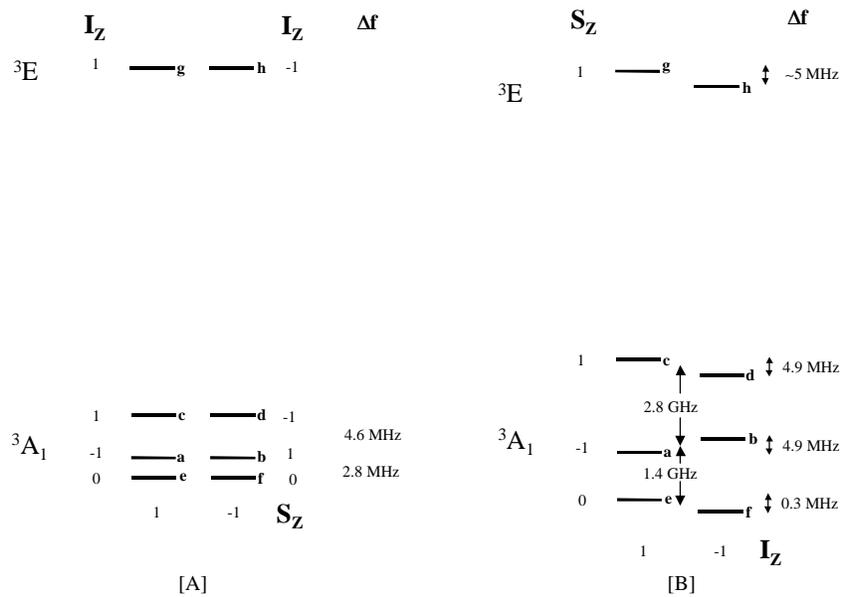

**FIGURE 4**